\pdfoutput=1
\documentclass{article}
\usepackage[margin=1in]{geometry}
\usepackage{booktabs}

\usepackage[final]{microtype}
\usepackage{subcaption}
\usepackage[x11colors]{xcolor}
\usepackage{tikz}
\usetikzlibrary{shapes,decorations.pathreplacing,decorations.pathmorphing,arrows,positioning,matrix,calc}
\tikzset{>=stealth}
\usepackage{amsmath,amssymb,amsthm,mathtools}
\usepackage[scr=boondox]{mathalfa}
\usepackage[colorlinks]{hyperref}

\usepackage{listings}

\tikzset{
	block/.style = {draw, rectangle, 
		minimum height=1cm, 
		minimum width=2cm},
	input/.style = {coordinate,node distance=1cm},
	output/.style = {coordinate,node distance=4cm},
	arrow/.style={draw, -latex,node distance=2cm},
	pinstyle/.style = {pin edge={latex-, black,node distance=2cm}},
	sum/.style = {draw, circle, node distance=1cm}
}

\numberwithin{equation}{section}

\theoremstyle{definition}

\theoremstyle{remark}

\begin{document}
\title{ Simulated Blockchains for Machine Learning Traceability and Transaction Values in the Monero Network }
\author{Nathan Borggren, Hyoung-yoon Kim, Lihan Yao and Gary Koplik}
\date{Geometric Data Analytics, Inc., Durham, NC, 27701}
\maketitle

\begin{abstract}
    
Monero is a popular crypto-currency which focuses on privacy.  
The blockchain uses cryptographic techniques to obscure transaction values as well as a `ring confidential transaction' which seeks to hide a real transaction among a variable number of spoofed transactions.  
We have developed training sets of simulated blockchains of 10 and 50 agents, for which we have control over the ground truth and keys, in order to test these claims.  
We featurize Monero transactions by characterizing the local structure of the public-facing blockchains and use labels obtained from the simulations to perform machine learning.
Machine Learning of our features on the simulated blockchain shows that the technique can be used to aide in identifying individuals and groups, although
it did not successfully reveal the hidden transaction values.  
We apply the technique on the real Monero blockchain to identify ShapeShift transactions, 
a cryptocurrency exchange that has leaked information through their API providing labels for themselves and their users.


\end{abstract}

\section{Introduction}
We have successfully applied machine learning (ML) in the past by combining features derived directly from blockchains with labels aggregated from off-chain sources \cite{deep_borggren_2017, ss_borggren_2018}.
Other researchers have used ML techniques as well to deanonymize exchanges \cite{Ranshous2017},  identify malware \cite{DBLP:journals/corr/abs-1906-07852}, and 
other institutions \cite{zola2019cascading}.
With the exception of \cite{ss_borggren_2018}, these analyses have largely focused on Bitcoin as the labels are easily sourced, albeit difficult to verify \cite{walletexplorer}.
However, other crypto-currencies have also greatly progressed and provide sufficient liquidity to be incorporated into
cross-currency mixing schemes \cite{ss_borggren_2018,Meiklejohn:DBLP:journals/corr/abs-1810-12786}.

Although many crypto-currencies are derivative of Bitcoin, originating in a fork of the code repository followed by some cosmetic changes,
some coins have adopted different basic assumptions and introduced substantially new codebases and cryptographic techniques to overcome some of the
privacy limitations inherent in Bitcoin.  
One such coin is Monero \cite{xmr_github}.
Monero has introduced some structural differences to Bitcoin that were sufficient to render our
Bitcoin-like ML machinery inapplicable.

In particular, the following features of the Monero blockchain obstruct our previous analysis.
\begin{itemize}
    \item Addresses are not included in the blockchain, obfuscating the recipient of a transaction.
    \item Transaction values are obfuscated
    \item Inputs to a transaction are combined with spoofed transactions (the RingCT), obfuscating the origin of a transaction.
\end{itemize}

Despite these improvements, Monero is still susceptible to some level of tracking
using heuristics \cite{DBLP:journals/corr/MillerMLN17} and understanding residual effects of hard-forks \cite{DBLP:journals/corr/abs-1812-02808}.

Can we regain an ML approach to the deanonymization of Monero?  
A few hurdles need to be overcome.  
First, featurizations of the Monero blockchain will be heavily topological in their nature; 
timestamps, number of inputs, size of RingCTs and the connectivity among transactions are the raw materials from which we can make features.
Second, the monerod wallet client, in comparison with the bitcoind wallet client, has removed a number of features
that are of use to blockchain analysts but not necessary for the day to day use of the coin.
Lastly, large repositories of labels of transactions are either unavailable, unreliable, or classified.

To overcome the first hurdle, in Sec. 3, we will quantify some topological aspects in the blockchain neighborhood of a given transaction.  
We have utilized \cite{xmr_onion}, the open source repository behind \url{xmrchain.net}, 
to overcome the second hurdle.  
To address the third hurdle, we have collected and verified labels aggregated from the ShapeShift API.
These labels allows us to deanonymize only one entity, ShapeShift, and while we do so in Sec. 5 to demonstrate the use of our features on the real monero blockchain, 
but for the sake of generality and additional confidence we have chosen another course which we hope will inspire other privacy analysis on Monero and other blockchains.

\begin{figure}[h]
	\centering
	\includegraphics[width=.8\linewidth]{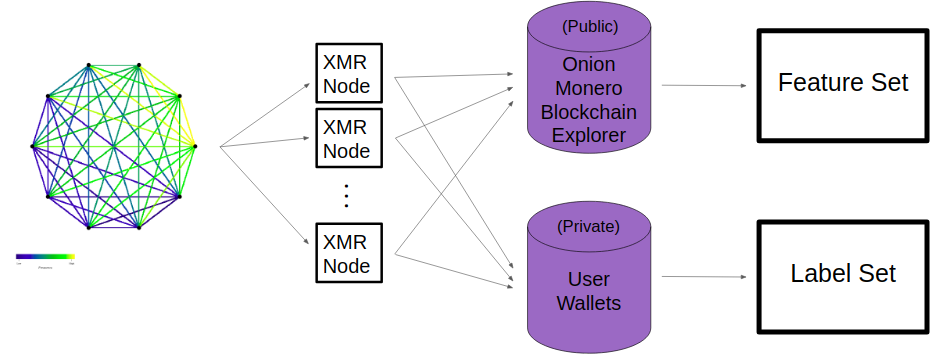}
	\caption{A test-network is used to generate a blockchain and provide labels.}	
	\label{fig:ml}
\end{figure}

In cybersecurity studies, it is commonplace to generate synthetic datasets to develop and assess threat detection techniques \cite{6565236}.
We use this analogy to provide a framework to understood the inner workings of Monero in action; 
test networks are created including ten and fifty person networks to provide a dataset for an ML analysis.
For the ML assessment, the blockchain, that is the public-facing aspect of the network, is used to develop featurizations 
while the wallet contents are used to supply the labels as depicted in fig \ref{fig:ml}.

Monero also provides new opportunities for machine learning that were not present or applicable in the Bitcoin-like case.  In Section 4, 
in addition to exploring the agent identification task we will explore these other use cases.  
For example, we can use our features in a regression analysis seeking to find the missing transaction value.  
Although this is shown to be unsuccessful in this case, we did achieve success in recovering the real input out of the spoofed transactions.

\section{Simulating Blockchains}

To simulate a Monero economy where multiple agents transact with each other and mine simultaneously, we use tmux, an application for creating and managing multiple terminal sessions on a single machine. 
We run a shell script to create a tmux-based network of Monero agents based on hard-coded wallet information.
On this network, each agent runs a mining operation and a wallet remote procedure call (RPC) attached to an allocated set of ports. 
All agents on the network are connected to each other and keep a synchronized blockchain.

The “economy files” specify how the economy will operate. 
Each economy file contains a list of transaction amounts, destination addresses, and wait times between the transactions that the particular agent will adhere to. 
We generate these files with a stochastic model. 
We use a Python program to process these files and make HTTP-based transaction requests to the Monero wallet RPCs accordingly. 
 
Other than the constraint that the wait times between transactions must be sufficient to keep the delays in transaction processing relatively small, the content of the economy files is up to imagination. 
We consider several economic scenarios with varying assumptions and create the economy files accordingly. 

Three of the five scenarios we consider are ten-agent economies, and the rest are fifty-agent economies. 
The wait intervals between transactions and the transaction amounts are generated by the Poisson distributions. 
For scenario “s03,” the agents’ wait intervals come from varying Poisson parameters ranging from 45 to 90,000, while the transaction amounts come from the same Poisson parameter for all agents. 
Scenario s04 uses varying Poisson parameters for generating transaction amounts as well. 
Scenario s05 is the same as s03, but the economy is divided into two pools. 
That is, the agents only transact within the pool they belong to. 
Telling these pools apart based on public transaction data is one of the machine learning questions we pursue. 

\begin{figure}\
	\centering
	\begin{subfigure}[h]{0.24\linewidth}
	\includegraphics[width=\linewidth]{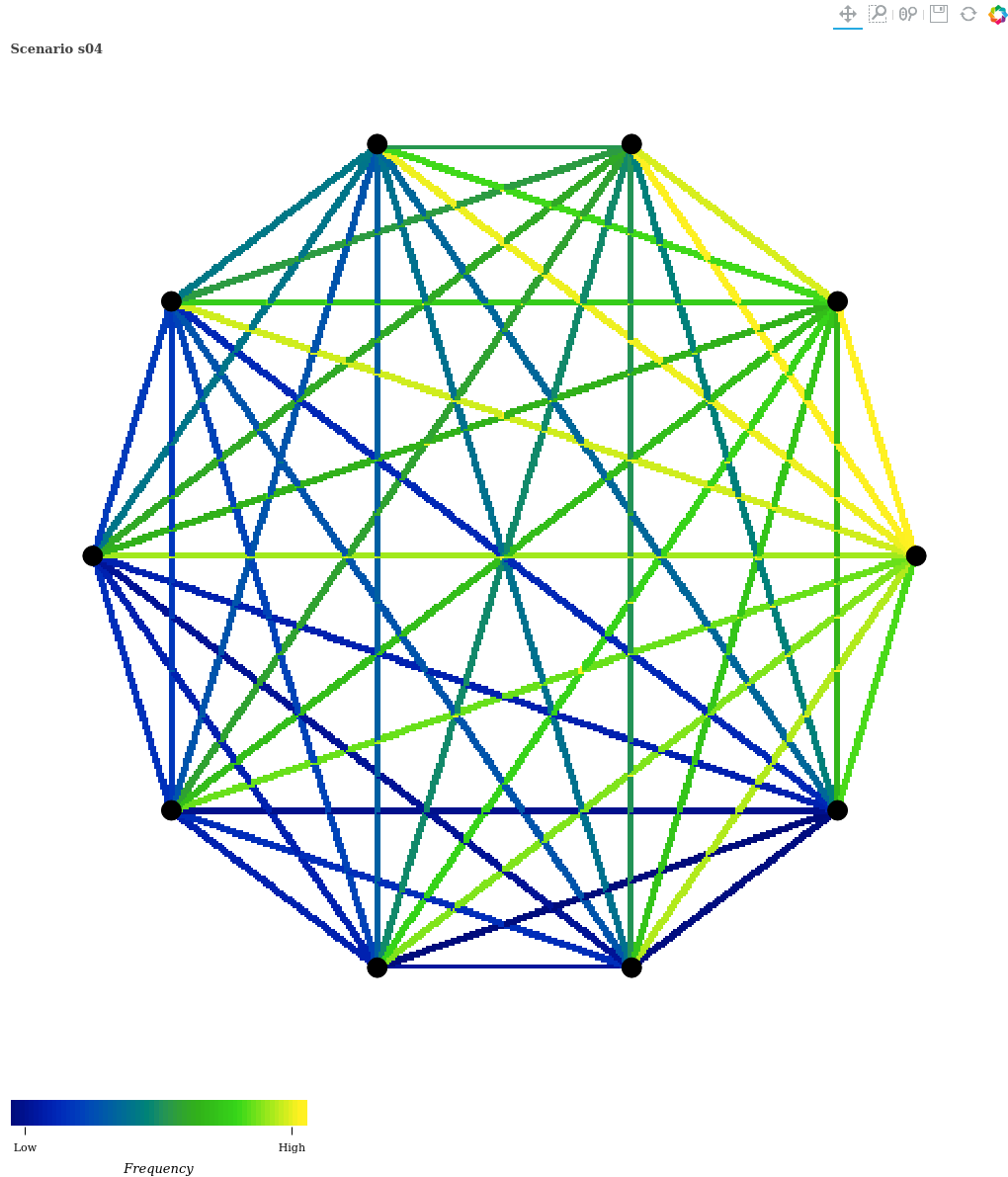}
	\caption{graph for s03 and s04 }
	\end{subfigure}
	\hfill
	\begin{subfigure}[h]{0.24\linewidth}
        \includegraphics[width=\linewidth]{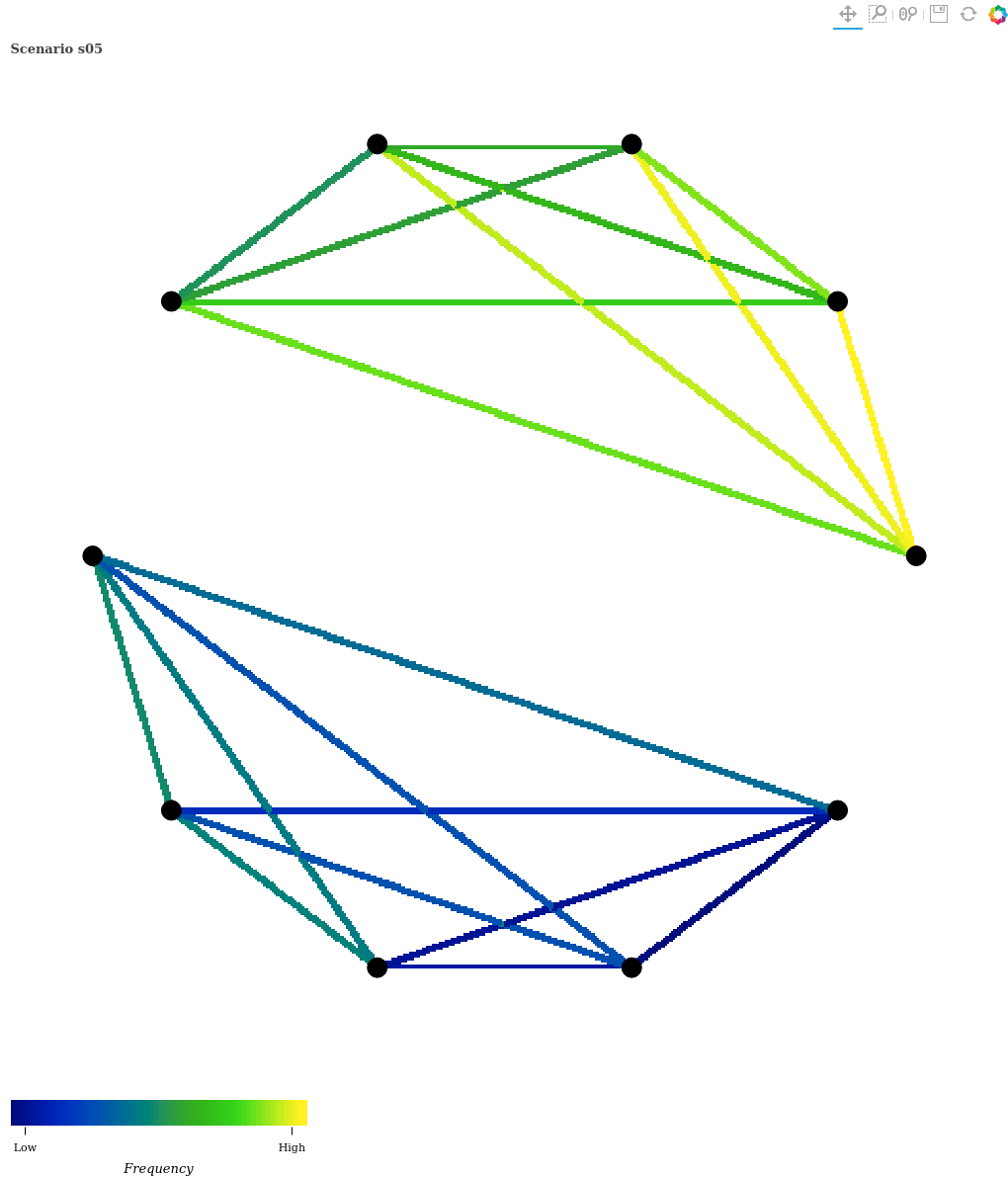}
        \caption{graph for s05 }
        \end{subfigure}
    \begin{subfigure}[h]{0.24\linewidth}
        \includegraphics[width=\linewidth]{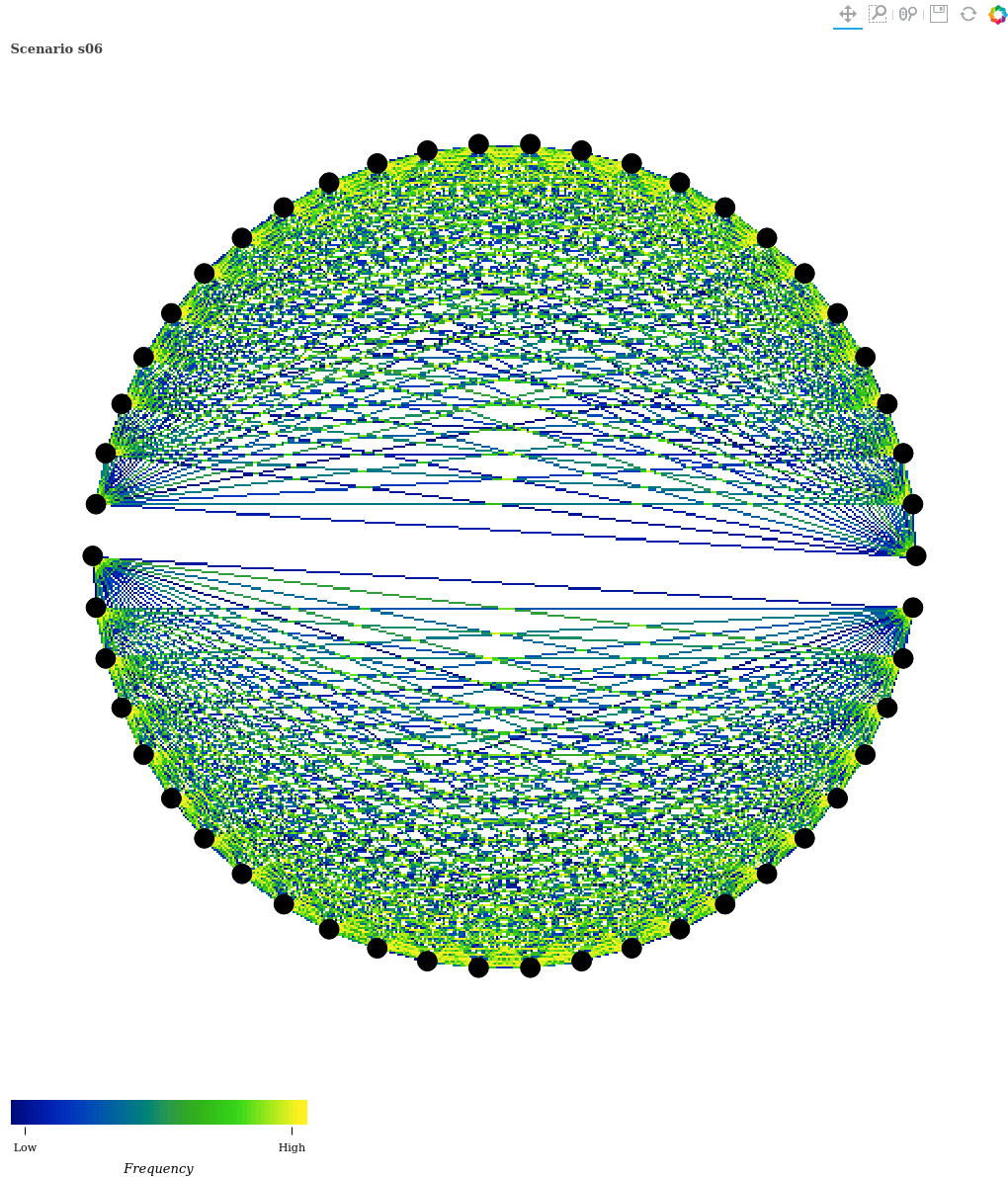}
        \caption{graph for s06 }
        \end{subfigure}
        \hfill
        \begin{subfigure}[h]{0.24\linewidth}
            \includegraphics[width=\linewidth]{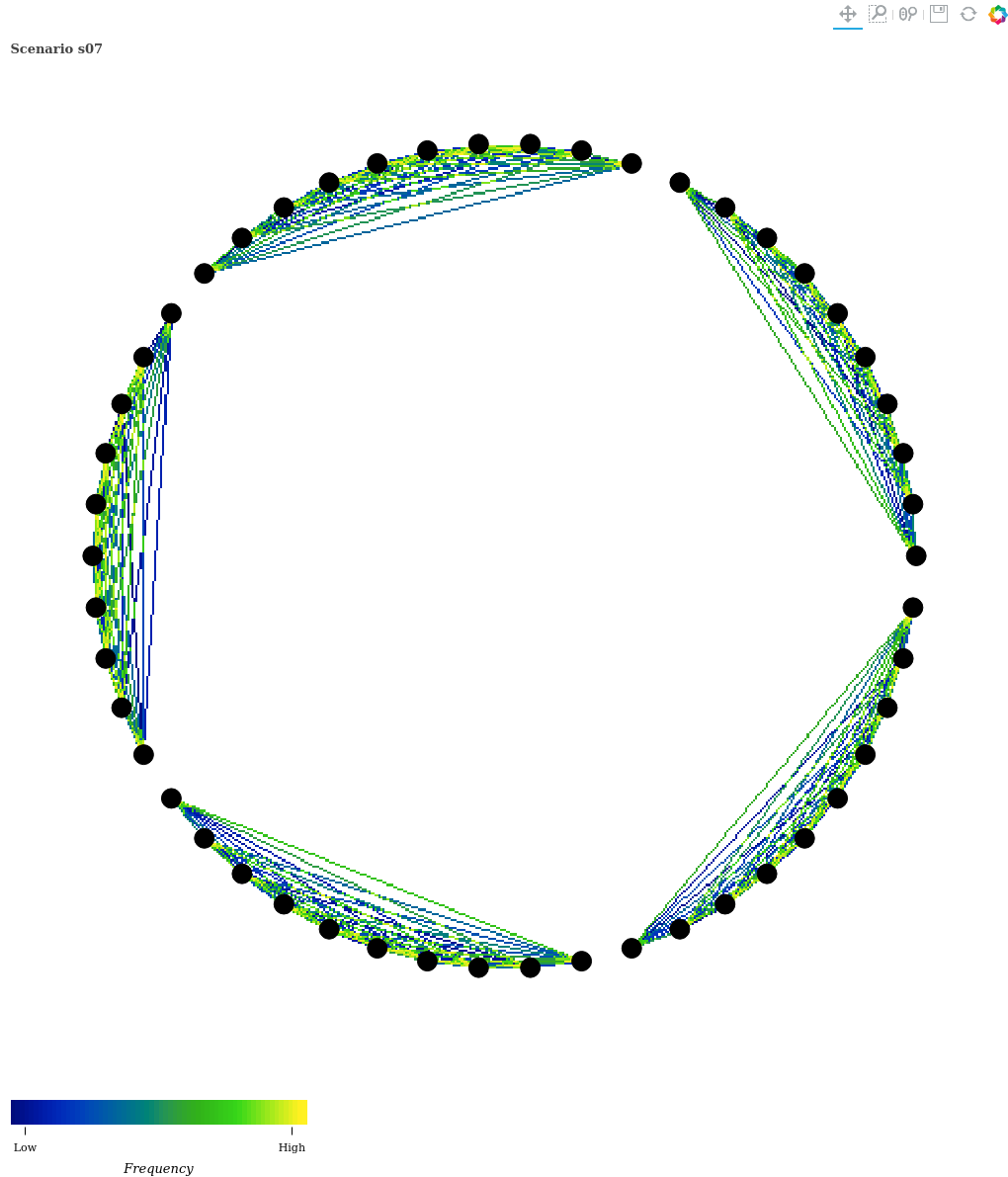}
            \caption{Graph and s07 }
            \end{subfigure}
    \caption{Graphs of economy files, showing connectivity of trading partners with edges, and color shows transaction size.}
	\label{fig:econ_files}
\end{figure}

For one of the two fifty-agent scenarios (s06), we create two pools of equal number of agents. 
Additionally, we create two different cycles of transactions for the two pools, simulating a time zone difference in the real world. 
Finally, in our final fifty-agent scenario (s07), we devise a transaction network with five ten-agent pools.  
Fig. \ref{fig:econ_files} shows the experimental setup.

\begin{figure}[ht]\
	\centering
	\begin{subfigure}[h]{0.49\linewidth}
	\includegraphics[width=\linewidth]{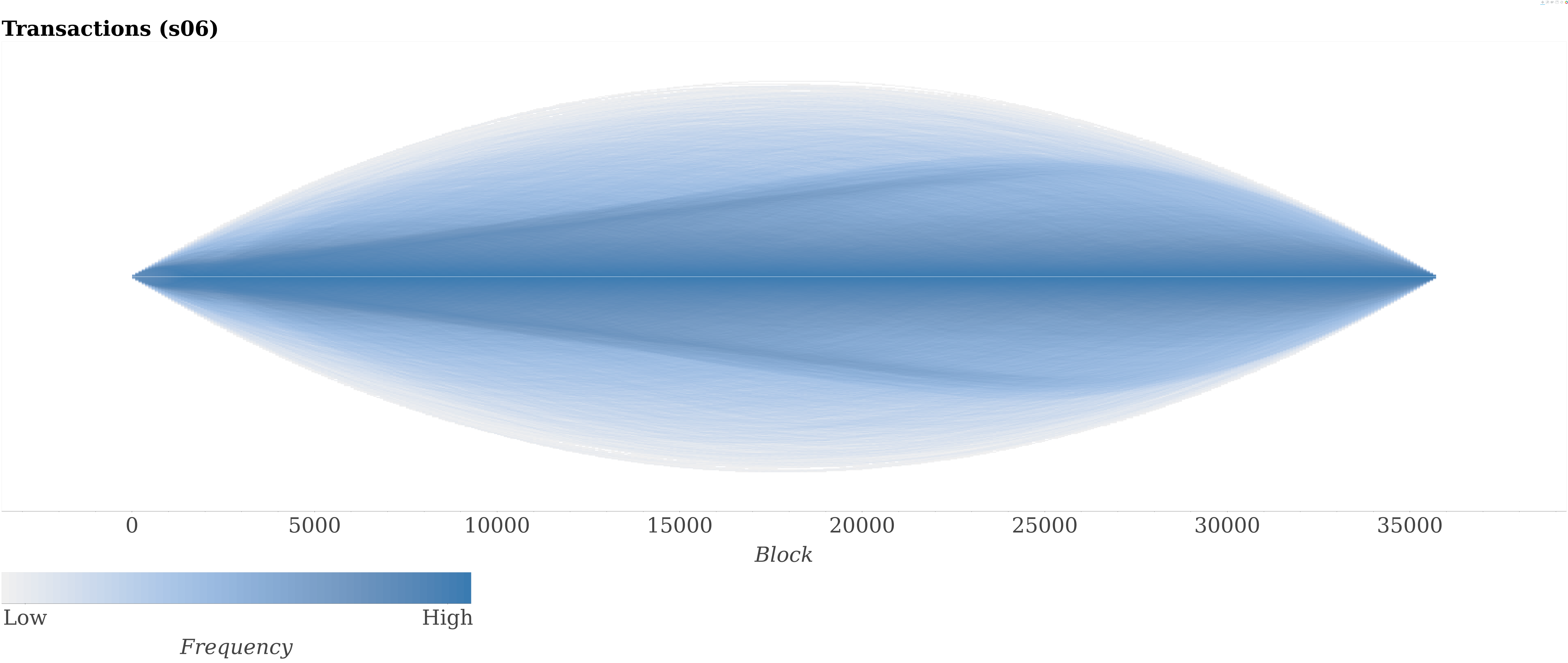}
	\caption{For s06, all edges between transactions are shown}
	\end{subfigure}
	\hfill
	\begin{subfigure}[h]{0.49\linewidth}
        \includegraphics[width=\linewidth]{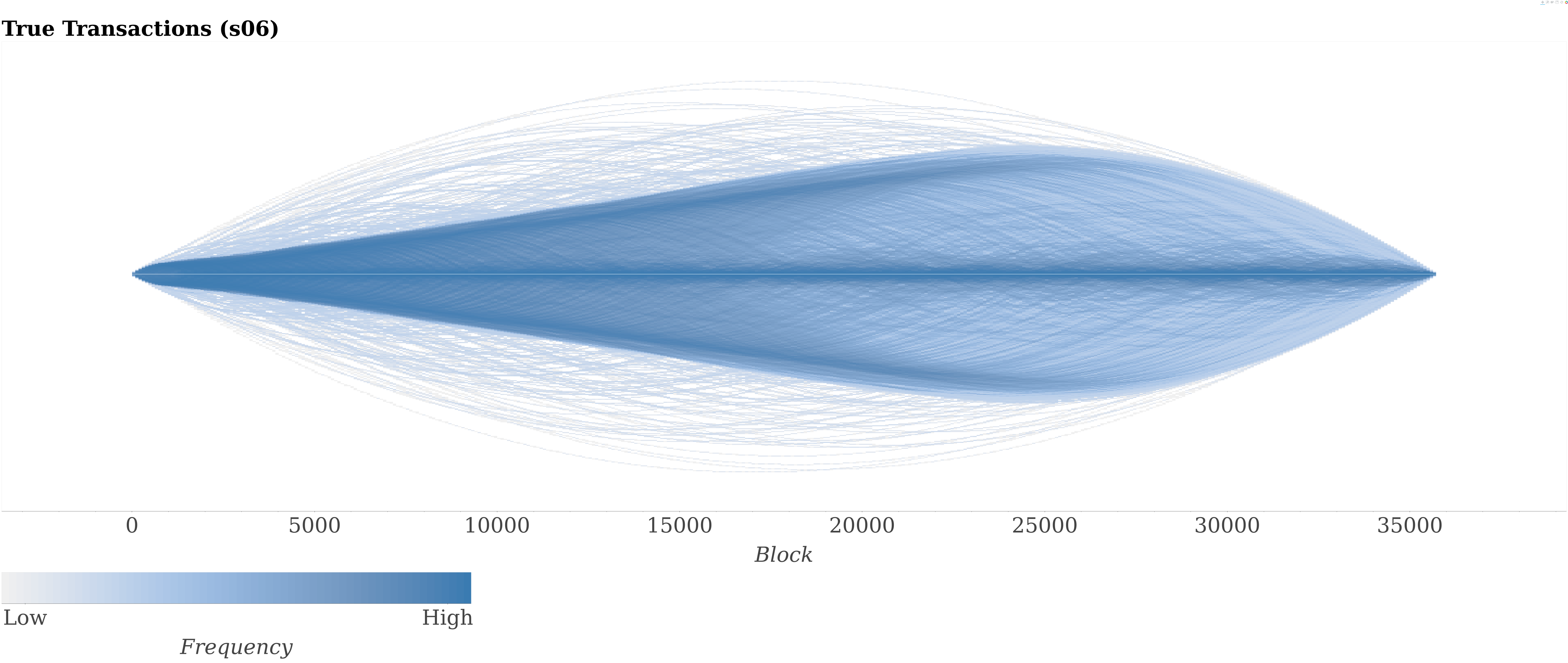}
        \caption{For s06, only true edges between transactions are shown}
	\end{subfigure}%
	\caption{Removing the spoofed transactions shows a much clearer image of the blockchain activity.}
	\label{fig:true}
\end{figure}

To examine the blockchain data, we link the Onion Monero Blockchain Explorer to the blockchain of one of the agents.
Table 1 shows a summary of the datasets generated, 
while fig \ref{fig:true} shows the dramatic sparsity that can be achieved by removing the spoofed transactions from the blockchain.

\begin{table}[h]
	\centering
\begin{tabular}{l|r|r}
	\toprule
	\textbf{Simulation} &  \textbf{Number of Blocks}   &   \textbf{Number of Transactions}  \\ \hline
	\midrule
	 \textbf{s03} &    23812 &       4898 \\\hline
	 \textbf{s04} &    25509 &        4923 \\\hline
    \textbf{s05} &    41583 &        4923 \\ \hline
	 \textbf{s06} &   37281 &        24807 \\ \hline
	 \textbf{s07}  & 58551 & 7070  \\ \hline
	\bottomrule
	\end{tabular}

	\caption{Summary of Simulation Datasets}
\end{table}
\label{tab:sims}

\section{Featurizing Monero Blockchains}
\paragraph{Real Monero Blockchain}

For most machine learning tasks on real cryptocurrency blockchains, 
the difficulty in procuring labels for supervised learning remains a central challenge, despite 
the accessibility of blockchain activities themselves. In Monero, this is especially so.  

In certain situations, labels and unseen blockchain information may be 
collected by law enforcement, hedge funds and hobbyists. We utilize transaction data from \href{https://shapeshift.io/#/coins}{Shapeshift service}.
The dataset contains around 1,700,000 transaction entries of which 20,000 are ShapeShift transactions being converted into Monero.   

For each transaction entry, the data contains 7 0-hop features - features intrinsic to
the transaction as it appears on the Monero explorer interface, e.g. transaction timestamp, ring size,
day of week, hour of day, etc. From the 0-hop features,
 175 1-hop features, aggregate statistics including mean, max, standard deviation, 
 regarding the transaction neighbors' 0-hop features are also collected. The 182 features are Z-normalized
 prior to the predictive task.

 \begin{figure}[h]
	\centering
	\includegraphics[width=.8\linewidth]{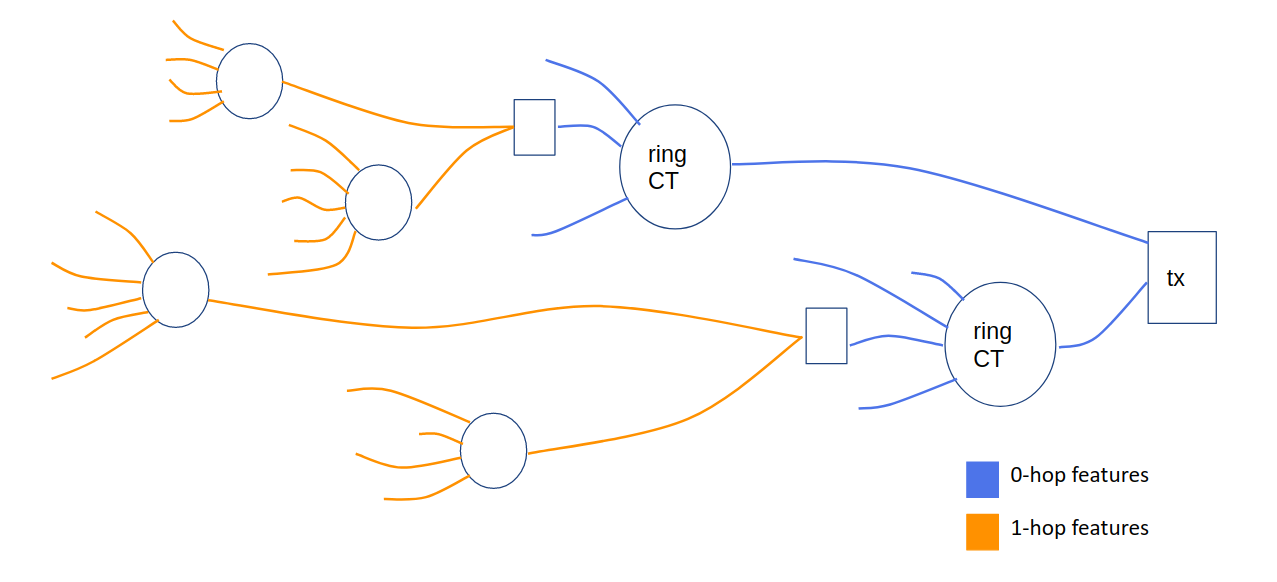}
	\caption{A neighborhood around a transaction is featurized by collecting statistics of the 0-hop and 1-hop transactions.}	
	\label{fig:ml_features}
\end{figure}

 This approach considers the neighborhood information of a particular transaction, while 
 not requiring the extraction of the entire network. We find that the 1-hop features 
 have been informative in Shapeshift transaction predictions. On test nets, 
 group membership prediction and transaction value regression tasks have also benefited from neighborhood information. 
 
In a related work \cite{borggren_multi}, we propose a correlation statistic for mitigating the noise introduced by the mixin mechanism.
This feature follows from the hypothesis regarding multiple wallet usage: when a 
transaction contains multiple ring CTs, the real inputs within each ring are contributed by the same user 
or users exhibiting similar behavior. For example, a receiver may expect a token amount exceeding the 
value stored within a single wallet, so the sender provides inputs from multiple wallets that were 
\emph{ previously traded at a similar frequency themselves}. We quantify this intuition by
tabulating a population statistic over transactions containing exactly two rings. The $(i,j)$-th 
entry of the correlation matrix is the correlation in time between $i$-th oldest entry of the first ring
with $j$-th oldest entry of the second. The binning of the matrix can be presented in hours of the day, or 
by partitioning the relative timestamp difference between inputs $i$, $j$.  
While our features do not explicitly reference the measured correlation matrix, 
our features have been chosen to be sensitive to these effects.

 \paragraph{ Test Nets } Of the five test nets s03, s04, s05, s06, s07, all were inputs to the 
 transaction value regression task, while group membership prediction was studied in the last three test nets. 
 The preprocessing of Monero explorer information and featurization follows the featurization process of the 
 real Monero blockchain.

\section{Machine Learning}
\subsection{Machine Learning Spoofed/Real Transactions}
An interesting new ML task that Monero uniquely offers is in identifying which of a given set of RingCT elements was the real previous transaction.
Our correlation analysis suggests that pattern-of-life behaviors, such as a users' typical transaction time of day can reveal information about the true input.
The features we have computed are sensitive to these differences and give reason to suggest that ML can recover real signal 
despite the mixins.

\begin{figure}[h]
	\centering
	\includegraphics[width=.8\linewidth]{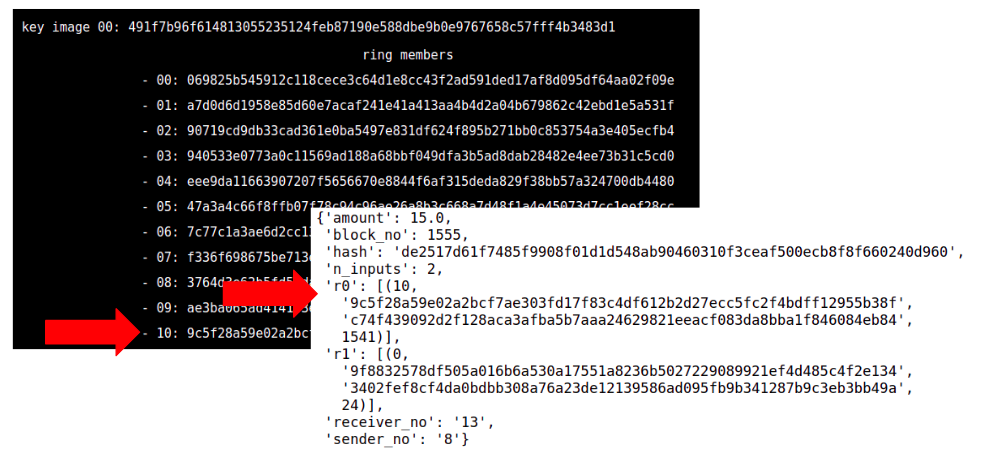}
    \caption{In black background is the public facing blockchain, in white are the known transactions of the given user. 
    The ML exercise is to recover the index of the real transaction.}	
	\label{fig:ml_spoof}
\end{figure}

Fig. \ref{fig:ml_spoof} shows a depiction of the task at hand. 
In black is a screenshot of the blockchain with an arrow in red showing the true input.  
The true input is known by analyzing the wallets, shown in white, of the users after the simulation terminates.

\subsection{Machine Learning User Identity and Group Membership}
In the task of identifying users and groups from transaction logs, we applied neural networks and 
random forest for classification. An observation in this case is a transaction - we classify the group that the 
transaction's receiver belongs to. The groups across scenarios trade strictly within-group, and only during designated 
time intervals. These time intervals repeat in a cyclic pattern, akin to timezones.

\begin{figure}[h]
	\centering
	\includegraphics[width=.8\linewidth]{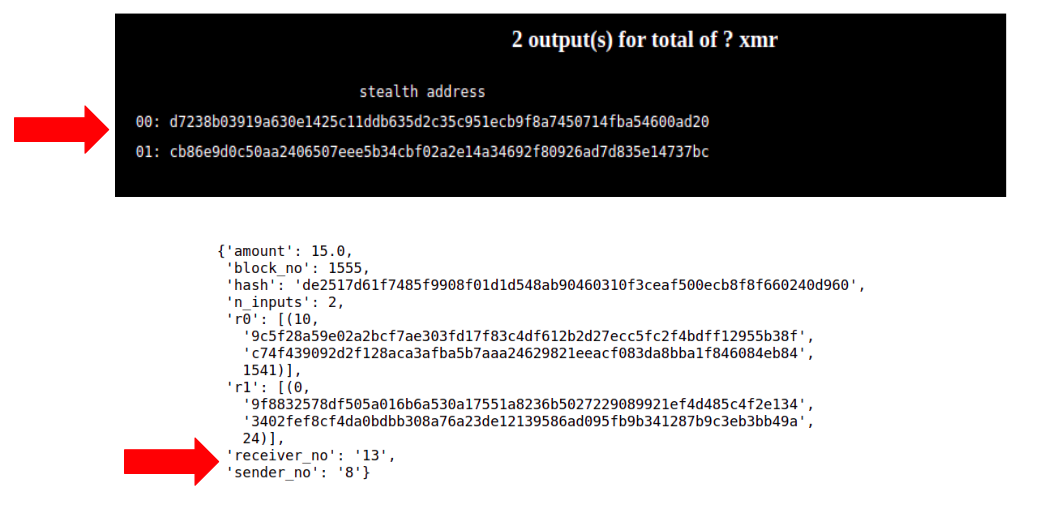}
	\caption{The membership in a group is known from the wallet files and economy instructions.}	
	\label{fig:ml_faction}
\end{figure}

For both models, a randomized search over hyperparameters was conducted. 
The neural network is composed of 2 dense layers with 182 and $n$ neurons respectively, where the second layer 
ranges from 10 to 30. 

The random search over random forest hyperparameters include: number of decision trees, maximum depth, 
maximum features, minimum samples per split, and the splitting criterion. Because random forest has 
an interpretable feature importance weighting, the top 3 informative features for group membership
classification are:    
{\bf S05} The informative S05 features involve the input minute among a transaction's inputs. 
They are listed in decreasing order of importance:  
Sum of average input minutes, average of max input minutes, 
and average of minimum input minutes. 
{\bf S06} The informative features are: sum of input seconds, average of maximum input hours, 
and sum of max input hours
{\bf S07} The informative features are: median of input minute sums, maximum of input minutes, 
and average of maximum input minutes. 


\subsection{Machine Learning Transaction Value}


Model performance for the value regression task is the $R^2$ coefficient. 
\[R^2 =  1 - \frac{\sum_i (y_i - \tilde{y_i})^2}{ \sum_i (y_i - \mu_y )^2 }   \]

where $\mu_y$ is the transaction value in expectation. We set the baseline predictor of this task to be 
predicting shapeshift transaction values in expectation, and its $R^2$ performance is 0. A perfect score 
is 1, and models that improve on our baseline performance lies in the range $(0, 1]$.  
 
To predict the transaction values of the i-th entry, we fitted two models: Epsilon-Support Vector 
Regression (SVR) and neural network. 

For both models, a randomized search was completed over hyperparameters. The SVR hyperparameters include its
 kernel, regularization parameter C, and tolerance parameter $\epsilon$. The neural network is 
 composed of 2 dense layers, with 182 and $n$ neurons respectively, $n$ ranging within [10, 30]. 
 The learning rate $\gamma$ was another hyperparameter searched over. Given a hyperparameter set, model performances are computed over 5-folds of the data then averaged.   

We find that more information regarding transactions is required to predict their values. The SVR model 
has a $R^2$ value of $-0.16$, and the neural network has a value of $-0.1$, both below the 
baseline performance of $0$. Future directions of value regression include 

\begin{enumerate}
    \item In the case of multiple rings, incorporating input-level correlations between rings.
    \item Integrating known user-level information to value prediction.  
\end{enumerate}

\begin{figure}[h]
	\centering
	\includegraphics[width=.8\linewidth]{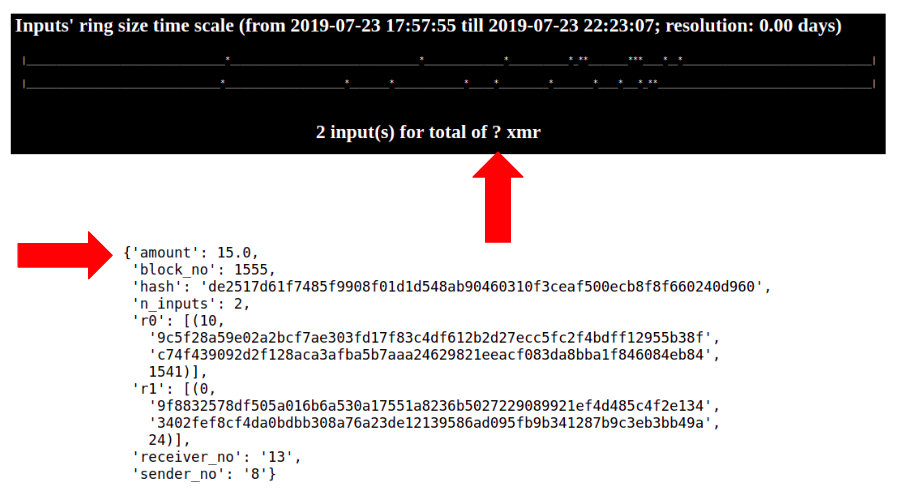}
    \caption{The transaction value on the blockchain is missing in the black. 
    We attempt to recover the value from the wallet in white through regression on the features.}	
	\label{fig:ml_value}
\end{figure}

\subsection{ML results}

\begin{figure}[h]
	\centering
	\includegraphics[width=.8\linewidth]{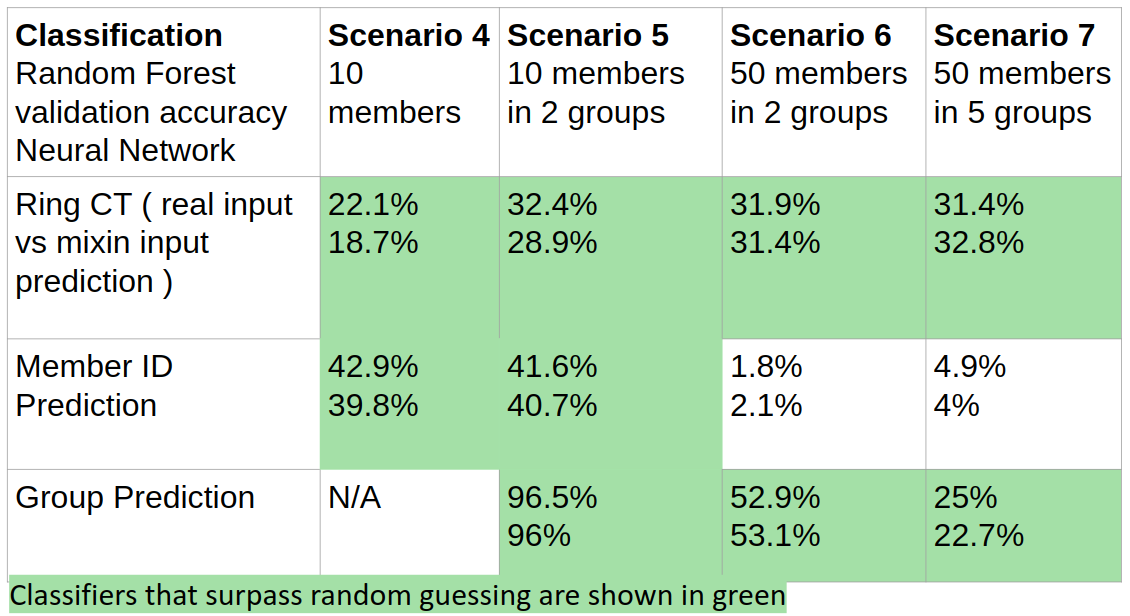}
    \caption{ML results for the various tasks are shown.  In transaction value predictions, our regressors did not surpass baseline of guessing average transaction values. }	
	\label{fig:ml_results}
\end{figure}

Fig. \ref{fig:ml_results} shows the results of our machine learning efforts.  
We expect that uniformity between the users and a severe lag in time transactions posted compared to the design specifications
contributed to the poor performance in Scenario 6 and Scenario 7.  

\section{Machine Learning on the real Monero network; Identifying ShapeShift}
\label{sec:transfer_real}
Previously in \cite{ss_borggren_2018} we had collected and validated ShapeShift transactions using their API.  
This collection of data provided labels for Bitcoin, ZCash, Litecoin, and Dash transactions.  
Our featurizations of Bitcoin-like blockchains allowed us to successfully recall $73\%$ of ShapeShift transactions while analyzing only $20\%$ of the Bitcoin blockchain.  

In Table 2, we have extended that analysis to identify ShapeShift transactions whose target currency was Monero.
Surprisingly, the classifier for Monero outperformed that for Bitcoin.  
We attribute this to the fact that during its peak in 2018, ShapeShift was responsible for upwards of $4\%$ of the entire Monero blockchain,
where only one in a thousand transactions in Bitcoin were attributed to ShapeShift.
Although our dataset was still greatly imbalanced, the extent of the imbalance was not nearly as severe.

\begin{table}[h]
       \centering
\begin{tabular}{l|l|r|r}
\toprule
\textbf{Metric} & \textbf{Summary Statistic} &  \textbf{ShapeShift}   &   \textbf{Not ShapeShift}  \\ \hline
\midrule
\textbf{Precision} & \textbf{Mean} &    0.050062 &        0.999161 \\
       & \textbf{SD} &    0.000129 &        0.000030 \\\hline
\textbf{Recall} & \textbf{Mean} &    0.941982 &        0.794504 \\ 
       & \textbf{SD} &    0.002058 &        0.000548 \\
\bottomrule
\end{tabular}
\caption{Our features enable a high recall of ShapeShift transactions, identifying $94\%$ of the transactions looking at $20\%$ of the blockchain.}
\end{table}

\begin{figure}[ht]\
	\begin{subfigure}[h]{0.65\linewidth}
	\includegraphics[width=\linewidth]{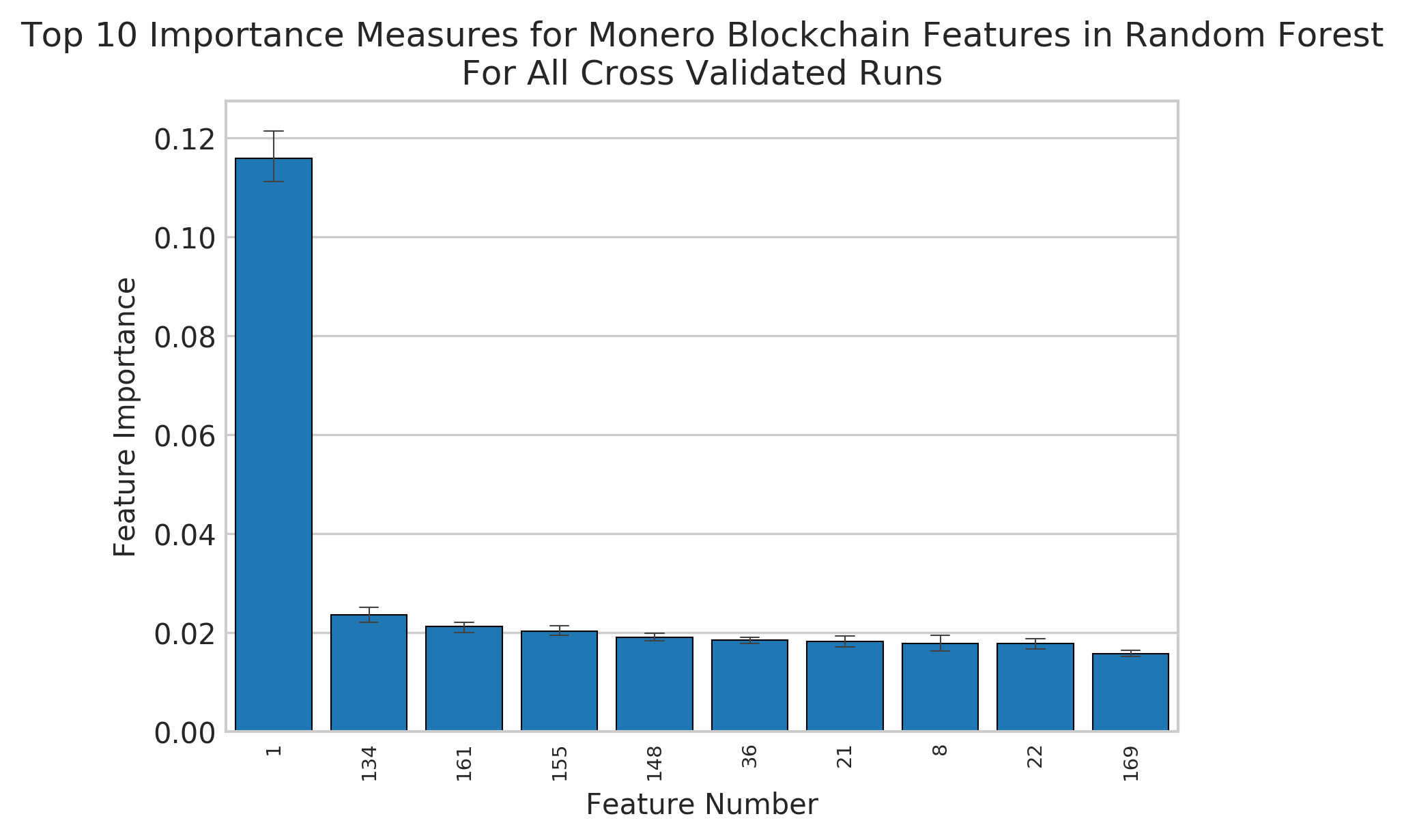}
	\caption{the features were ranked in accordance to their importance}
	\end{subfigure}
	\hfill
	\begin{subfigure}[h]{0.29\linewidth}
        \includegraphics[width=\linewidth]{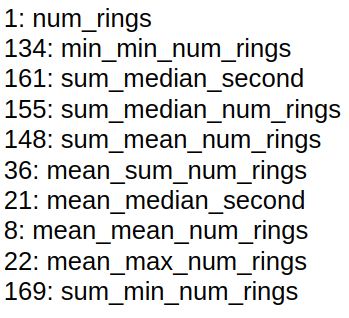}
        \caption{Descriptors for the important features}
	\end{subfigure}%
	\caption{Characteristics of the RingCT topology were largely informative.}
	\label{fig:featimport}
\end{figure}

The most important feature was by far the number of rings used in the tx, with seven of the other nine top-ten-features also involving summary statistics of the number of rings.
These are shown in Fig. \ref{fig:featimport}.

\section{Conclusions}
We have found that despite the complexities created by the enhanced privacy features of Monero, 
the blockchain is still susceptible to deanonymization and information gathering by machine learning techniques.
To establish this fact, test networks were developed to generate simulated blockchains.  
These blockchains were featurized while the wallets were parsed to provide labels.  
Monero proved to be robust against our efforts in recovery of the obfuscated transaction values, while classifiers for 
spoofed transaction identification and faction/group/user identification proved informative.

Additionally we have shown that ML on the real Monero blockchain can be used to identify a given party provided a sufficient collection of labels.
We used as our example the cryptocurrency exchange ShapeShift to demonstrate this, but expect this to hold true if other large collections of labels
were recovered, for example from wallets recovered in criminal investigations.
Despite having values for ShapeShift transactions, our regression attempts to recover these values proved unsuccessful.  
We hypothesize that future efforts could use much deeper features such as tracing all the way back to coinbase transactions, 
noting which version of Monero was being used for previous transactions, and incorporate features that look forward in time rather than the 
exclusively backwards features we have used.  
Such features may perhaps allow value information to leak into the features and provide success to a regression analysis, 
but the task would indeed be computationally expensive.

It is noted that recent versions of Monero now enforce the RingCT size to be eleven; as ring number derived features had been the most informative features
in the ShapeShift analysis, we expect this change to greatly enhance privacy going forward.  
However, as previous studies have shown, the residue from history of the past transactions will likely linger for some time as these 
new characteristics reshape the underlying distributions of features. 

We would like to thank the Machine Learning and Emerging Technologies exemplars at the LAS at North Carolina State University for funding and continuous feedback.
The irony does not escape us that the generosity of code and data from Monero, Onion explorer and the ShapeShift API, 
all participants devoted to privacy and security in the blockchain realm, were essential to this anti-privacy investigation and we are grateful for their contribution.






\newpage
\bibliography{bitcoin}

\begin{thebibliography}{10}

\bibitem{deep_borggren_2017}
Nathan Borggren.
\newblock Deep learning of entity behavior in the bitcoin economy.
\newblock
  \url{https://ncsu-las.org/wp-content/uploads/2017/12/borggren-gda_bitcoin.pdf},
  2017.

\bibitem{ss_borggren_2018}
Nathan Borggren, Gary Koplik, Paul Bendich, and John Harer.
\newblock Deanonymizing shapeshift: Linking transactions across multiple
  blockchains.
\newblock
  \url{https://ncsu-las.org/wp-content/uploads/2019/01/LAS_Shapeshift_Poster_1543186217.pdf},
  2017.

\bibitem{Ranshous2017}
Stephen Ranshous, Cliff~A. Joslyn, Sean Kreyling, Kathleen Nowak, Nagiza~F.
  Samatova, Curtis~L. West, and Samuel Winters.
\newblock {Exchange pattern mining in the bitcoin transaction directed
  hypergraph}.
\newblock {\em Lecture Notes in Computer Science (including subseries Lecture
  Notes in Artificial Intelligence and Lecture Notes in Bioinformatics)}, 10323
  LNCS:248--263, 2017.

\bibitem{DBLP:journals/corr/abs-1906-07852}
Cuneyt~Gurcan Akcora, Yitao Li, Yulia~R. Gel, and Murat Kantarcioglu.
\newblock Bitcoinheist: Topological data analysis for ransomware detection on
  the bitcoin blockchain.
\newblock {\em CoRR}, abs/1906.07852, 2019.

\bibitem{zola2019cascading}
Francesco Zola, Maria Eguimendia, Jan~Lukas Bruse, and Raul~Orduna Urrutia.
\newblock Cascading machine learning to attack bitcoin anonymity, 2019.

\bibitem{walletexplorer}
Ales Janda.
\newblock Wallet explorer.
\newblock \url{https://www.walletexplorer.com}, 2013-2017.

\bibitem{Meiklejohn:DBLP:journals/corr/abs-1810-12786}
Haaroon Yousaf, George Kappos, and Sarah Meiklejohn.
\newblock Tracing transactions across cryptocurrency ledgers.
\newblock {\em CoRR}, abs/1810.12786, 2018.

\bibitem{xmr_github}
fluffypony et~al.
\newblock Monero project.
\newblock
  \url{https://github.com/monero-project/monero/blob/master/src/wallet/wallet2.cpp},
  2015.

\bibitem{DBLP:journals/corr/MillerMLN17}
Andrew Miller, Malte M{\"{o}}ser, Kevin Lee, and Arvind Narayanan.
\newblock An empirical analysis of linkability in the monero blockchain.
\newblock {\em CoRR}, abs/1704.04299, 2017.

\bibitem{DBLP:journals/corr/abs-1812-02808}
Abraham Hinteregger and Bernhard Haslhofer.
\newblock An empirical analysis of monero cross-chain traceability.
\newblock {\em CoRR}, abs/1812.02808, 2018.

\bibitem{xmr_onion}
moneroexamples et~al.
\newblock onion-monero-blockchain-explorer.
\newblock
  \url{https://github.com/moneroexamples/onion-monero-blockchain-explorer},
  2016.

\bibitem{6565236}
J.~{Glasser} and B.~{Lindauer}.
\newblock Bridging the gap: A pragmatic approach to generating insider threat
  data.
\newblock In {\em 2013 IEEE Security and Privacy Workshops}, pages 98--104, May
  2013.

\bibitem{borggren_multi}
Nathan Borggren and Lihan Yao.
\newblock Correlations of multi-input monero transactions, 2019.

\end{thebibliography}
\bibliographystyle{unsrt}

\end{document}